\documentclass[twocolumn,showpacs,pra,aps,groupedaddress]{revtex4-1}

\usepackage{amsmath}
\usepackage{amsfonts}
\usepackage{amssymb}
\usepackage{graphicx}
\usepackage{hyperref}
\usepackage{bm}
\usepackage{color}
\usepackage{multirow}

\usepackage{times}
\usepackage{dsfont}
\usepackage{siunitx} %pour faire des jolis micro
\usepackage{etoolbox} % for \appto
\usepackage{lipsum} % for mock text
\usepackage[capitalize]{cleveref}
\hypersetup{backref,
colorlinks=true,
linkcolor=blue,
linktoc=black,
citecolor=cyan,
urlcolor=black}
\usepackage{extarrows}

\newcommand{\kb}{\ensuremath{{k}_\text{B}}}

\newcolumntype{P}[1]{>{\centering\arraybackslash}p{#1}}
 
\newcommand{\pc}[1]{\ensuremath{\left(#1\right)}} 
\newcommand{\px}[1]{\ensuremath{\left\lbrace#1\right\rbrace}} 
 
\newcommand{\ket}[1]{\ensuremath{\left\vert#1\right\rangle}}

\definecolor{burgundy}{rgb}{0.5, 0.0, 0.13}
\definecolor{denim}{rgb}{0.08, 0.38, 0.74}
\definecolor{midnightgreen}{rgb}{0.0, 0.29, 0.33}
\definecolor{sienna}{rgb}{0.53, 0.18, 0.09}
\definecolor{sacramentostategreen}{rgb}{0.0, 0.34, 0.25}

\newcommand\xyplane{$x$\nobreakdash--$y$~plane}

\makeatletter
\appto{\appendix}{%
  \@ifstar{\def\theequation@prefix{A.}}%
          {}%
}
\makeatother

\DeclareSIUnit\gauss{G}

\begin{document}
\title{Loss features in ultracold $^{162}$~Dy gases: two- versus three-body processes}

\author{Maxime Lecomte}
\author{Alexandre Journeaux}
\author{Loan Renaud}
\author{Jean Dalibard}
\author{Raphael Lopes}
\email{raphael.lopes@lkb.ens.fr}

\affiliation{
Laboratoire Kastler Brossel, Coll\`ege de France, CNRS, ENS-PSL University,
Sorbonne Universit\'e, 11 Place Marcelin Berthelot, 75005 Paris, France}

\begin{abstract}
\noindent Dipolar gases like erbium and dysprosium have a dense spectrum of resonant loss features associated with their strong anisotropic interaction potential. These resonances display various behaviours with density and temperature, implying diverse microscopic properties. Here, we quantitatively investigate the low-field ($B < \SI{6}{\gauss}$) loss features in ultracold thermal samples of $^{162}$Dy, revealing two- and three-body dominated loss processes. We investigate their temperature dependence and detect a feature compatible with a $d$-wave Fano-Feshbach resonance, which has not been observed before. We also analyse the expansion of the dipolar Bose-Einstein condensate as a function of the magnetic field and interpret the changes in size close to the resonances with a variation in the scattering length. 
\end{abstract}

\maketitle

%%%%%%%%%%%%%%%%%%%%%%%%%%%%%%%%%%%%%%%%%%%%%%
%%%%%%%%%%%%%%%%%%%%%%%%%%%%%%%%%%%%%%%%%%%%%%
%%%%%%%%%%%%%%%%%%%%%%%%%%%%%%%%%%%%%%%%%%%%%%
%%%%%%%%%%%%%%%%%%%%%%%%%%%%%%%%%%%%%%%%%%%%%%

\section{Introduction}

Quantum gases are by definition relatively short-lived, as these systems are extremely sensitive to loss processes such as collisions with residual gases, photo-association or inelastic collisions. Three-body losses, for example, correspond to inelastic recombination in which three particles interact sufficiently strongly to form a two-body bound state (dimer), while the third particle ensures energy conservation by acquiring a kinetic energy equal to the potential energy difference. This energy is usually much greater than the trap depth, resulting in the effective loss of all three particles. Such a mechanism is enhanced close to a scattering resonance. In inhomogeneous gases, three-body losses are particularly damaging as they lead to the depletion of the denser part of the atomic cloud, resulting in anti evaporative cooling \cite{Rem2013, Eismann2016}. However, while this process limits the timescales over which ultracold dense systems can be studied, it also provides an insight into the few-body physics of strongly interacting cold gases, which remains a challenging and stimulating area of research \cite{Chevy2016, Greene2017, Li2018,Yudkin2019,Ji2022, Yudkin2023}.
 
Dipolar gases like chromium, dysprosium, erbium, and thulium possess a large dipolar magnetic moment, resulting in properties that markedly differ from those of alkali atoms. 
In these systems, long-range anisotropic dipolar interactions lead to new features of the collision potential, such as the emergence of a $1/R^4$ potential \cite{Bohn2009} or the modification of the van der Waals $C_6$ coefficient \cite{Kotochigova2014}. This anisotropic dipolar interaction can lead to striking new behaviours, such as low-temperature $d$-wave Fano-Feshbach resonances \cite{Beaufils2009}. Furthermore, the anisotropic interaction potential is responsible for the dense spectrum of loss resonances  \footnote{ In here, we refer to these resonances as loss features instead of Fano-Feshbach resonances to allow for the possibility that some loss features are not associated with a change in the two-body scattering length but instead a genuine three-body effect.} in ultracold gases of erbium, dysprosium and thulium \cite{Baumann2014,Frisch2014,Maier2015,Khlebnikov2019}. A precise characterization of loss features has recently regained interest, triggered by the determination of the temperature dependence of their chaotic statistics \cite{Khlebnikov2019,Khlebnikov2021}, the optimization of the evaporative cooling \cite{Krstajic2023}, and the identification of appropriate Fano-Feshbach resonances in dipolar mixtures \cite{Durastante2020}.

In this article, we investigate the few-body processes driving the large number of low-field loss features in ultracold gases of $^{162}$~Dy. We recover the 9 previously reported resonances for this isotope \cite{Baumann2014} as well as 10 extra features, and quantitatively characterize the dependence on atom number and temperature for 11 features, indicated by the red vertical lines in Fig.~\ref{fig:fig1}. In addition, we identify a loss feature compatible with a $d$-wave Fano-Feshbach resonance, with similar characteristics to those reported for chromium in Ref.~\cite{Beaufils2009}. We also measure the three-body loss rate parameter for a Bose-Einstein condensate near zero magnetic field. This low-field zone is especially interesting if the objective is to establish spin-orbit coupling in a dysprosium gas, similar to \cite{Chalopin2020}, while  preventing two-body spin relaxation at increased atomic densities. Finally, we complement our analysis with a study of the BEC expansion near the different loss features, interpreting the dilatation of the cloud as a signature of enhanced two-body interactions.

%%%%%%%%%%%%%%%%%%%%%%%%%%%%%%%%%%%%%%%%%%%%%%
%%%%%%%%%%%%%%%%%%%%%%%%%%%%%%%%%%%%%%%%%%%%%%
%%%%%%%%%%%%%%%%%%%%%%%%%%%%%%%%%%%%%%%%%%%%%%
%%%%%%%%%%%%%%%%%%%%%%%%%%%%%%%%%%%%%%%%%%%%%%

\section{Low-field loss features}
\label{sec:exp1}

\begin{figure}[t!]
    \centering
    \includegraphics{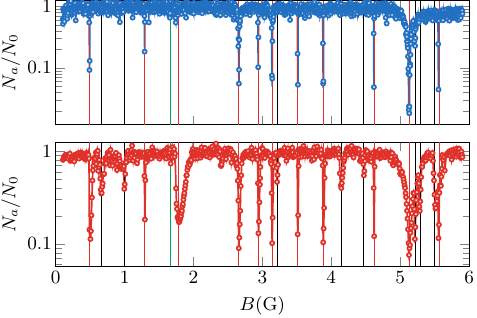}
    \caption{ Normalized atom number as a function of magnetic field for a thermal sample with temperature $T = \SI{190}{\nano\kelvin}$ (top panel in blue) and $T = \SI{2.4}{\micro\kelvin}$ (bottom panel in red). The red vertical lines indicate the 11 loss features for which we characterize the density and temperature dependence.}
    \label{fig:fig1}
\end{figure}

Experimentally, we load about $4\times 10^6$ dysprosium atoms ($ ^{162}$Dy) into a crossed dipole trap formed by laser beams operating at a wavelength $\lambda = \SI{1064}{\nano\meter}$. The atoms are loaded from a five-beam compressed MOT with atom number $N_a \approx 1 \times 10^8$ and temperature $T\approx \SI{15}{\micro \kelvin}$ (see Sec.~\ref{sec:app2} for more details). 

After forced evaporative cooling, we produce an ultracold thermal gas, spin polarized in the Zeeman sublevel of lowest energy $\ket{J, m_J = -J}$, with atom number $N_a=2 \times 10^5$ and temperature $T\sim \SI{190}{\nano\kelvin}$, above the condensation threshold. The evaporation is carried out with a fixed magnetic field of \SI{1.660}{\gauss}, indicated by the green vertical line in Fig.\ref{fig:fig1}. We then quench the magnetic field to a target value and hold the cloud for \SI{2}{\second}. We measure the atom number and temperature after time-of-flight absorption imaging, which gives us information about both the losses and heating of the cloud. The magnetic field is scanned from 0 to \SI{6}{\gauss} and we measure the atom number variation as a function of the magnetic field $B$. For this low temperature, we identify 10 loss features, corresponding to the atom number drops in Fig.~\ref{fig:fig1} top panel. These resonances, except for one, have been reported in Ref.~\cite{Baumann2014}. 

A similar experiment is performed for a hotter thermal cloud with temperature $T = \SI{2.4}{\micro\kelvin}$. We recover the previous resonances and observe several new loss features (lower panel in Fig.~\ref{fig:fig1}), increasing the total number of loss features count to 19. This result qualitatively demonstrates the non-trivial emergence of temperature-dependent loss features in an ultracold gas of dysprosium \cite{Baumann2014}.

To probe different temperatures $T$, we use a protocol that differs from most previous investigations of resonant losses in lanthanides, by preparing all of our samples in the same optical potential regardless of $T$. We do this by (i) cooling the atoms to very low temperature ($\sim 200$\,nK), (ii) adiabatically recompressing the optical trap to a large depth, and (iii) tuning the temperature by parametric heating of the trapped gas. Before measuring the remaining atom number, we ensure that the cloud is in thermal equilibrium by holding it for \SI{0.5}{\second}, which is long compared to the elastic collision time. By contrast, many previous studies adjusted temperature by halting the evaporation process at varying laser intensities. Our protocol addresses a potential bias resulting from differences in polarizability between free atoms and the bound state involved in the loss process \cite{Khlebnikov2021}. 
Additionally, it is advantageous to operate at a high trap depth $U_0$, leading to a large ratio $\eta = U_0 / \kb T$. This effectively reduces the losses due to evaporation that could potentially obscure distinctive features resulting from the loss resonances we seek to examine.

%%%%%%%%%%%%%%%%%%%%%%%%%%%%%%%%%%%%%%%%%%%%%%
%%%%%%%%%%%%%%%%%%%%%%%%%%%%%%%%%%%%%%%%%%%%%%
%%%%%%%%%%%%%%%%%%%%%%%%%%%%%%%%%%%%%%%%%%%%%%
%%%%%%%%%%%%%%%%%%%%%%%%%%%%%%%%%%%%%%%%%%%%%%

\section{Microscopic description of loss features} \label{sec:theory}

Before continuing with our experimental analysis, let us summarize the models developed in Ref.~\cite{Beaufils2009, Maier2015, Li2018, Waseem2018} that have been used to characterize loss features in dipolar gases. These models assume as an intermediate step the resonant formation of a dimer or a trimer, and they lead to different scaling laws of the maximal loss rate with density, as we show now. To keep the analysis simple, we assume here a uniform atomic density $n_a$, but the following results can then be readily transposed to the case of a harmonically trapped gas.

%%%%%%%%%%%%%%%%%%%%%%%%%%%%%%%%%%%%%%%%%%%%%%
%%%%%%%%%%%%%%%%%%%%%%%%%%%%%%%%%%%%%%%%%%%%%%

\subsection{The resonant dimer model}
\label{subsec:dimer_th}

\begin{figure}[t!]
    \centering
    \includegraphics{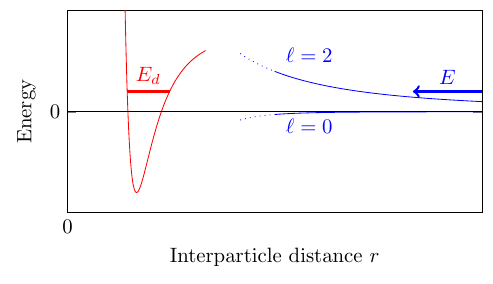}
    \caption{Resonant dimer model. A pair of atoms, with relative motion of energy $E$ and angular momentum $\ell$, can resonantly form a dimer state of energy $E_d\approx E$, which then decays at a rate $\Gamma_d$. In the case of a narrow Fano-Feshbach resonance \cite{Chin2010}, this process leads to a sharp energy feature described by the Lorentzian of Eq.~\ref{eq:Lorentzian}.}
    \label{fig:figdimer}
\end{figure}

We describe a sequential process involving two steps (i) a quasi-resonant coupling between a state with two free atoms and a state where a dimer in some excited state $A_2^*$ is formed:
\begin{equation}
A+A \rightleftarrows	A_2^*\, ,
\label{eq:A_plus_A}
\end{equation} 
(ii) the decay of the dimer with a rate $\Gamma_d$. Here, this  decay is essentially induced by the collision with a third atom: $ A_2^* + A \rightarrow A_2 + A$ so that $\Gamma_d$ is implicitly a function of $n_a$, and where $A_2$ is a deeply bound dimer. 

In the following we restrict ourselves to the case of large $\Gamma_d$, so that the dimer $A_2^*$ decays soon after its formation by the direct process in (\ref{eq:A_plus_A}), and the reverse process in (\ref{eq:A_plus_A}) does not play a significant role. This scenario can happen in the case of a narrow Fano-Feshbach resonance and a sufficiently large atomic density $n_a$.  We will see that it is the relevant one for most of the resonances observed in our experimental conditions and we refer the reader to  Refs. \cite{Beaufils2009,Li2018,Waseem2018,Waseem2019} for a discussion of the general case and in the limit of small $\Gamma_d$. Note that the situation considered here is the opposite of that of Ref.~\cite{Zhang2023}, where the atoms $A$ formed a Bose-Einstein condensate and where the authors could observe a coherent oscillation between the two members of Eq.~\ref{eq:A_plus_A} thanks to bosonic stimulation. 

To model the process (\ref{eq:A_plus_A}), we consider a fictitious box containing two atoms, whose typical volume is thus $L^3=2/n_a$, where $n_a$ is the atomic density. We work in the center-of-mass frame of the two atoms and we denote $E$ the energy of the relative motion of the colliding atoms. We suppose that the resonant process (\ref{eq:A_plus_A}) occurs for an incident partial wave $\ell$ and we introduce  the coupling matrix element $\hbar \kappa$ between the relative wave function of the two free atoms and the wave function of the dimer state,
of energy $E_d$ (see Fig.~\ref{fig:figdimer}). Each wave function is supposed to be normalized to unity in the box $L^3$.  
We will not try to provide here a detailed expression for $\kappa$ and we simply note its scaling with energy and box size: $\kappa \propto (E^\ell/L^3)^{1/2}$ \cite{Landau2013}. The hamiltonian of this fictitious two-level system is thus the $2\times 2$ matrix
\begin{equation}
\begin{pmatrix}
E & \hbar \kappa  \\ \hbar \kappa^*  &  E_d-i\hbar\Gamma_d/2
\end{pmatrix}
\end{equation}
where we have added the imaginary term $-i\hbar\Gamma_d/2$ to the dimer energy to account for its instability. The two eigenvalues of this matrix are complex, which expresses the fact that the scattering state $A+A$ is now also unstable because of its coupling to $A_2^*$. 

The imaginary part of the energy of the $A+A$ pair in the presence of the coupling $\kappa$ gives the decay rate of this pair, which reads for $\Gamma_d\gg \kappa$ (Breit-Wigner formula):
\begin{equation}
\Gamma_a \approx \frac{\Gamma_d |\kappa|^2}{(E-E_d)^2/\hbar^2+\Gamma_d^2/4} \, ,
\label{eq:Breit_Wigner}
\end{equation}
from which we deduce the scaling of the total loss rate in the sample with $N_a$ atoms at a given energy  $E$:
\begin{equation}
\dot N_a\propto -N_a \Gamma_a \propto - n_a N_a E^\ell {\cal L}(E-E_d)
\label{eq:dNa_dt}
\end{equation} 
where ${\cal L}$ stands for the Lorentzian function:
\begin{equation}
{\cal L}(x)=\frac{1}{2\pi}\frac{\hbar\Gamma_d }{x^2+(\hbar \Gamma_d/2)^2}.
\label{eq:Lorentzian}
\end{equation}

We now average the rate (\ref{eq:dNa_dt}) over a thermal distribution of temperature $T$, so that the loss rate $\dot N_a$  is proportional to
\begin{equation}
\frac{\int_0^{+\infty} dE\;\rho(E)\;E^\ell\; {\cal L}(E-E_d) \;e^{-E/k_{\rm B}T}  }{
\int_0^{+\infty} dE\;\rho(E) \;e^{-E/k_{\rm B}T} 
}\, ,
\label{eq:thermal_average}
\end{equation}
with the density of states $\rho(E)\propto \sqrt E$ for the relative motion of the $A+A$ pair.  
The expression (\ref{eq:thermal_average}) is in general a complicated function of $E_d$ and $\Gamma_d$ (and thus $n_a$). An interesting limiting case is obtained when the resonance width $\hbar\Gamma_d$ is very small compared to $k_{\rm B}T$, in which case the Lorentzian function ${\cal L}(E-E_d)$ can be replaced by a Dirac function $\delta(E-E_d)$ in (\ref{eq:thermal_average}). In this  case, we find that the decay rate corresponds to a two-body loss process \cite{Beaufils2009}:
\begin{equation}
\dot N_a=-L_2 n_a N_a
\label{eq:dot_Na_2}
\end{equation}
with an effective two-body loss parameter $L_2$ given by:
\begin{equation}
L_2(E_d,T) \propto T^{-3/2}\;E_d^{\ell+1/2}\;e^{-E_d/k_{\rm B}T}.
\label{eq:loss_rate_L2}
\end{equation}

The experimental procedure involves scanning the value of $E_d$ by ramping the magnetic field at a given temperature and searching for the maximal loss rate. From the scaling of Eq.~(\ref{eq:loss_rate_L2}), we find that the maximum loss rate occurs for $E_d=(\ell+1/2)k_{\rm B}T$ with
\begin{equation}
L_2^{\rm (max)}(T)\propto (k_B T)^{\ell-1}.
\label{eq:L2_max}
\end{equation} 
Assuming the two hypotheses above are valid, i.e. $\hbar \kappa\ll \hbar\Gamma_d\ll k_B T$, the variation of   $L_2^{\rm (max)}$ with temperature thus gives immediate access to the partial wave $\ell$ involved in the resonant loss process. The validity of these hypotheses can be checked by verifying that $\dot N_a/N_a$ scales linearly with $n_a$ (see Eq.(\ref{eq:dot_Na_2})).

%%%%%%%%%%%%%%%%%%%%%%%%%%%%%%%%%%%%%%%%%%%%%%
%%%%%%%%%%%%%%%%%%%%%%%%%%%%%%%%%%%%%%%%%%%%%%

\subsection{The resonant trimer model}
\label{subsec:trimer_th}

The second model  consists in a pure three-body process \cite{Maier2015}. Three particles that do not have resonant pairwise interactions arrive through a three-body open channel $\mathcal{O}_1$, with the quantum number $\lambda$ associated with the grand angular momentum, and an incoming energy $E$ close to the energy $E_t$ of an excited trimer state $A_3^*$, residing in a closed channel $\mathcal{C}$: 
\begin{equation}
A+A+A \rightleftarrows A_3^* \, .
\end{equation}
If we neglect atom interactions at long distance, the incoming channel is purely repulsive, even for $\lambda=0$, with the 3-body centrifugal potential $V(R) \propto \left[\lambda(\lambda+4) + \frac{15}{4}\right]/R^2$, where $R$ is the hypergeometric radius \cite{Delves1960, Smith1960}. The closed channel $\mathcal{C}$ is coupled to other channels $\mathcal{O}_f$ that are not directly coupled to the incoming open channel, thereby determining the decay rate $\Gamma_t$ of the trimer $A_3^*$. Note that since we do not assume resonant two-body interactions, the trimer $A_3^*$ differs from Efimov trimers. The latter play an important role in broad Fano-Feshbach resonances and lead to the $1/T^2$ dependence of the three-body loss rate $L_3$ \cite{Rem2013,Petrov2015}. 

The analysis of this process follows along the same general lines as for the resonant dimer model.
We consider a fictitious box of volume $L^3\sim 3/n_a$ containing three particles. The coupling $\kappa$ between the incoming state of energy $E$ and the resonant trimer state  now scales  as $\kappa\propto E^{\lambda/2}/L^3$ and the width $\Gamma_t$ of the trimer state $A_3^*$ induces a non-zero width for the incoming state $A+A+A$, given by a formula similar to (\ref{eq:Breit_Wigner}). We are then led to $\dot N_a=-n_a^2 N_a E^\lambda {\cal L}(E-E_t)$ where $\cal L$ is the Lorentzian function similar to (\ref{eq:Lorentzian}) with width $\Gamma_t$ \cite{Dincao2005b,Wang2011,Maier2015} \footnote{Note however that the scaling in $E^\lambda$ has been questioned for non-zero values of $\lambda$ by \cite{Wang2011} for some specific cases.}. The thermal average of this decay rate involves the three-particle density-of-state $\rho(E)\propto E^2$ so that we obtain, in the limit $\hbar \Gamma_t\ll k_{\rm B}T$:
\begin{equation}
\dot N_a=-L_3n_a^2 N_a
\label{eq:dot_Na_3}
\end{equation}
with the scaling $L_3\propto T^{-3} \,E_t^{\lambda+2}\,e^{-E_t/k_{\rm B}T}$. When scanning the energy of $E_t$ by ramping the external magnetic field, the maximum loss rate scales as
\begin{equation}
L_3^{\rm max}(T)\propto (k_{\rm B}T)^{\lambda-1}
\label{eq:L3_max}
\end{equation}
and it is obtained for $E_t=(\lambda+2)\,k_{\rm B}T$.

\paragraph*{Discussion.} Although the scaling of the loss rate with temperature is similar in both models of \S\ref{subsec:dimer_th} and \S\ref{subsec:trimer_th}, the latter is a pure three-body process and therefore cannot predict two-body dominated features of the type $\dot N_a/N_a\propto -n_a$. The reverse statement may not be true: Refs.~\cite{Beaufils2009,Li2018, Waseem2019} indicate that the resonant dimer model can lead to an effective three-body loss rate $\dot N_a/N_a\propto -n_a^2$ in the case where the excited bound state is long-lived relative to its coupling to the incoming open channel. Let us also emphasize that the simple scaling laws (\ref{eq:L2_max},\ref{eq:L3_max}) hold only when  $\hbar\kappa \ll \hbar\Gamma_{d,t}\ll k_{\rm B}T$. If this is not the case, the variation of $L_{2,3}^{\rm (max)}$ with temperature is non-trivial.

%%%%%%%%%%%%%%%%%%%%%%%%%%%%%%%%%%%%%%%%%%%%%%
%%%%%%%%%%%%%%%%%%%%%%%%%%%%%%%%%%%%%%%%%%%%%%
%%%%%%%%%%%%%%%%%%%%%%%%%%%%%%%%%%%%%%%%%%%%%%
%%%%%%%%%%%%%%%%%%%%%%%%%%%%%%%%%%%%%%%%%%%%%%

\section{Losses vs. density and temperature}

In this section, we present our experimental results regarding the loss rates in a harmonically trapped $^{162}$Dy gas due to inelastic processes. We first outline our methodology (\S\ref{subsec:methodology}), and then investigate the variations of the loss rate with density at a given temperature (\S\ref{subsec:density_variation}), and  with temperature  at a given atom number (\S\ref{subsec:T_variation}). We summarize our results for the whole set of resonances in \S\ref{subsec:all_features}.

%%%%%%%%%%%%%%%%%%%%%%%%%%%%%%%%%%%%%%%%%%%%%%
%%%%%%%%%%%%%%%%%%%%%%%%%%%%%%%%%%%%%%%%%%%%%%

\subsection{Methodology}
\label{subsec:methodology}

We recall that the microscopic nature of the process (2-body or 3-body loss) of each resonance is unknown, as is the temperature variation of the associated loss rate ($L_2(T)$ or $L_3(T)$). Therefore, for a gas prepared with a given atom number $N_a$ and temperature $T$, we restrict our analysis of the decay rate to a short time interval $\Delta t$, during which $N_a$ and $T$ vary by less than 20\% and 30\%, respectively. A linear fit $N_a(t) = N_0 (1- \beta t)$ to the decaying atom number over this interval then allows us to derive the decay rate $\dot N_a\approx \Delta N_a/\Delta t$ for a density $\bar n_a$ and a temperature $\bar T$ taken equal to the average value of these quantities over the time interval $\Delta t$. The interval retained for the fit is indicated by a coloured zone in Figs. \ref{fig:losses_vs_n} and \ref{fig:losses_vs_T}.
 
In addition, we recall that the volume of a  trapped gas in a harmonic potential is a function of the temperature. If interactions play a negligible role, one finds 
$V = (2\sqrt{3} \pi \kb T/m \bar{\omega}^2)^{3/2}$,
where $\bar{\omega} = \pc{\omega_x \omega_y \omega_z}^{1/3}$ is the geometric mean of the trapping frequencies, and $\kb$ the Boltzmann constant. In the following, we designate by $n_a=N_a/V$ the average density in the trap. For a fixed temperature, the volume does not vary, and $\beta=-\dot N_a/N_a$ scales as $\beta \propto n_a^\gamma$ with $\gamma =1$ (resp. $\gamma=2$) in the case of a two-body (resp. three-body) dominated loss process (\ref{eq:dot_Na_2}) [resp. (\ref{eq:dot_Na_3})]. 
 
%%%%%%%%%%%%%%%%%%%%%%%%%%%%%%%%%%%%%%%%%%%%%%
%%%%%%%%%%%%%%%%%%%%%%%%%%%%%%%%%%%%%%%%%%%%%%

\subsection{Examples at fixed $T$: 2- versus 3-body dominated losses}
\label{subsec:density_variation}

%%%%%%%%%%%%%%%%%%%%%%%%%%%%%%%%%%%%%%%%%%%%%%
\begin{figure}[t!]
    \centering
    \includegraphics{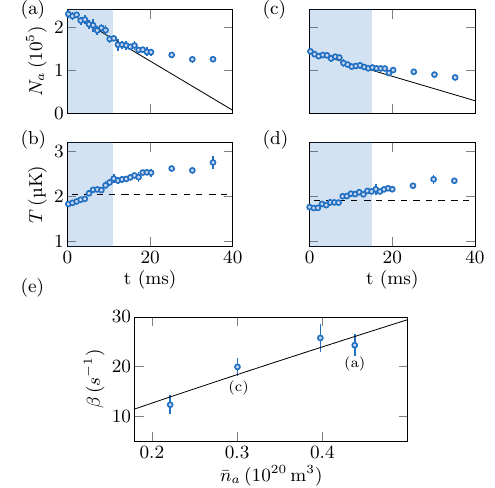}
    \caption{Atom loss dynamics for varying total atom number $N_a$ at a magnetic field $B=\SI{5.130}{\gauss}$. The averaged temperature is constant and equal to $T \approx \SI{2.0(1)}{\micro\kelvin}$. Panels (a) and (b) show the atom number and temperature evolution for the case of an initial atom number $N_0 = 2.3 \times 10^5 $. Panels (c) and (d) show the atom number and temperature evolution for $N_0 = 1.4 \times 10^5 $. (e) Variation of the initial loss rate, $\beta=-\dot N_a/N_a$, with the density $\bar{n}_a \propto \bar{N}_a/\bar{T}^{3/2}$. The solid line is the fitting function $\beta \propto \bar{n}_a^\gamma$ with $\gamma = 0.92(10)$.}
    \label{fig:losses_vs_n}
\end{figure}
%%%%%%%%%%%%%%%%%%%%%%%%%%%%%%%%%%%%%%%%%%%%%%

Here we describe a typical analysis procedure of one of the resonances featured in Fig. \ref{fig:fig1}, specifically the resonance occurring at a magnetic field strength of \SI{5.130}{\gauss}. As explained in \S~\ref{sec:exp1}, we prepare a sample with an adjustable atom number at the desired temperature. The depth of the trap is sufficient to render losses due to evaporation insignificant \footnote{We have independently estimated the lifetime of our vacuum system to be $\tau \gtrsim \SI{15}{\second}$.}. 
We jump the magnetic field from its initial value ($B=\SI{1.660}{\gauss}$) to a magnetic field close to the target loss feature (typically \SI{100}{} - \SI{200}{\milli\gauss} away from it). We then wait for \SI{500}{\milli\second} and perform a second quench towards the magnetic field for which the losses are maximum. This sequence allows a better resolution of the initial loss dynamics.

We show in Fig.~\ref{fig:losses_vs_n}a,c two decay curves $N_a(t)$ for two initial  atom numbers, hence two atomic densities $n_a$. Figs.~\ref{fig:losses_vs_n}b,d show the corresponding changes in temperature. A linear fit of the short time variation of $N_a$ provides the decay rate $\beta$ introduced in \S\,\ref{subsec:methodology}. We summarize our results for $\beta$ as a function of  $ n_a$ in Fig.~\ref{fig:losses_vs_n}e. A fit $\beta( n_a)=\beta_1 n_a^\gamma$  gives $\gamma = 0.92(10)$, an indication of a two-body dominated loss feature for this particular resonance. 

To cross-validate our methodology, we have performed a similar loss measurement for a thermal gas at $B=\SI{1.660}{\gauss}$ i.e. away from any loss resonance. We measure $\gamma = 2.3(4)$, which is consistent with a three-body loss process,  as expected for a gas with positive background scattering length. We find, for a thermal gas, $L_3 = 1.2(2) \times 10^{-40} \SI{}{\meter^6/\second}$.

The same technique allows us to determine the three-body loss coefficient of a Bose-Einstein condensate (BEC) either close to $B=0$ or at $B= \SI{1.660}{\gauss}$ away from any loss resonance. For that purpose we evaporate until we produce a quasi-pure BEC, and then adiabatically recompress the trap to a final trap depth of \SI{2.16}{\micro\kelvin}, with frequencies equal to $\px{\omega_x, \, \omega_y,\, \omega_z} = 2\pi \times \px{49,\, 152,\, 115} \SI{}{\hertz}$. 
We find $L_3 = 2.9(3) \times 10^{-41}\SI{}{\meter^3/\second}$ and $L_3 = 5.0(8) \times 10^{-41}\SI{}{\meter^3/\second}$, for the background three-body loss rate, $B=\SI{1.660}{\gauss}$, and the near-zero field $B\lesssim \SI{50}{\milli\gauss}$, respectively. The values of $L_3$ are extracted for a BEC fraction going from near unity to 0.5. Compared to a thermal sample, we find a reduction of $L_3$ in qualitative agreement with the predicted $3!$ reduction for a pure BEC \cite{Kagan1985}. 

%%%%%%%%%%%%%%%%%%%%%%%%%%%%%%%%%%%%%%%%%%%%%%
%%%%%%%%%%%%%%%%%%%%%%%%%%%%%%%%%%%%%%%%%%%%%%

\subsection{Example at fixed $N_a$: $T$-dependent loss rate}
\label{subsec:T_variation}

%%%%%%%%%%%%%%%%%%%%%%%%%%%%%%%%%%%%%%%%%%%%%%
\begin{figure}[t!]
    \centering
    \includegraphics{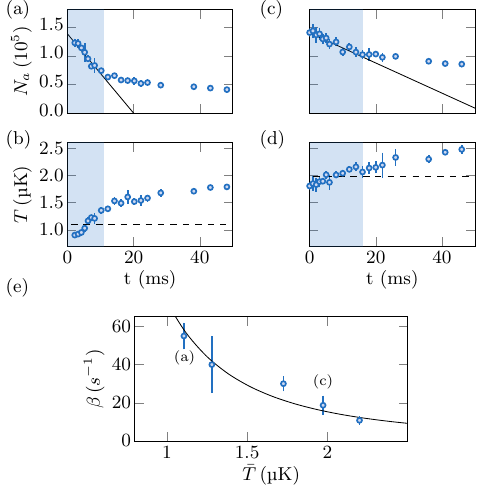}
    \caption{Atom losses as a function of temperature $T$ for a fixed atom number $N_a\approx 1.5\times 10^5$, for the loss feature $B = \SI{5.130}{\gauss}$. (a) and (b) Atom number and temperature evolution, respectively, for a thermal cloud with temperature $\bar T = \SI{1.1 (1)}{\micro\kelvin}$. (c) and (d) Atom number and temperature evolution for $\bar T = \SI{2.0(1)}{\micro\kelvin}$. (e) Variation of the initial loss rate, $\beta =-\dot N/N$, with temperature. The solid line is the fitting function $\beta \propto T^\alpha$, with $\alpha=-2.2(3)$. }
    \label{fig:losses_vs_T}
\end{figure}
%%%%%%%%%%%%%%%%%%%%%%%%%%%%%%%%%%%%%%%%%%%%%%

We now turn to the temperature dependence of the atom loss rate, still taking the resonance at $B=\SI{5.130}{\gauss}$ as an example. Following the procedure outlined in \S\ref{sec:exp1} and \S\ref{subsec:density_variation}, we prepare thermal samples at different temperatures but with the same atom number. We plot in Fig.~\ref{fig:losses_vs_T}a (resp. Fig.~\ref{fig:losses_vs_T}c) the atom number decay for the initial temperature $T = \SI{0.9}{\micro\kelvin}$ (resp. $T = \SI{1.8}{\micro\kelvin}$). We show in Fig.~\ref{fig:losses_vs_T}b,d the corresponding time evolution of the temperature. 

As explained in \S\,\ref{subsec:methodology}, we restrict to the short time evolution and extract the rate $\beta=-\dot N_a/N_a$ from a linear fit to the measured decay of $N_a(t)$. The values of $\beta$ for different temperatures are shown in Fig.~\ref{fig:losses_vs_T}e. We fit the relation $\beta(T) =\beta_2 T^\alpha$ to the temperature dependence of the loss rate, with $\alpha=-2.2(3)$. From the analysis of \S\ref{subsec:density_variation}, we know that this particular resonance is likely to be due to a two-body loss decay $\dot N_a=-L_2 n_a N_a$, hence $\beta\propto L_2/V$ for a given $N_a$. Recalling that the volume in a harmonic trap scales as $V\propto T^{3/2}$, we infer that $L_2(T)\propto T^{\chi_2}$ with $\chi_2=\alpha+3/2\approx -0.7(3)$ for this particular resonance.  

%%%%%%%%%%%%%%%%%%%%%%%%%%%%%%%%%%%%%%%%%%%%%%
%%%%%%%%%%%%%%%%%%%%%%%%%%%%%%%%%%%%%%%%%%%%%%

\subsection{Analysis of all loss resonances}
\label{subsec:all_features}

\begin{figure}[t!]
    \centering
    \includegraphics{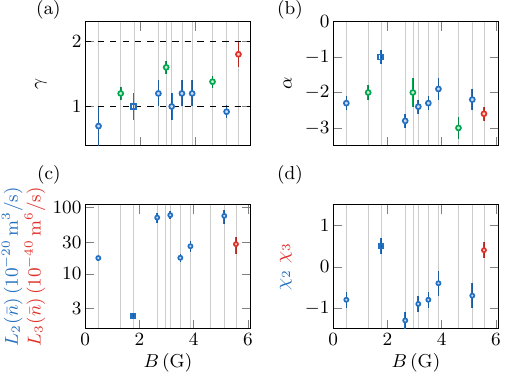}
    \caption{ Summary of density- and temperature-dependence for 11 loss features between 0-\SI{6}{\gauss}. (a) Determination of two- versus three-body dominated loss features. A value of $\gamma$ compatible with $1$ indicates a two-body dominated loss feature, while a value of $\gamma =2$ indicates a three-body dominated feature. These values are represented by dashed horizontal lines. Loss features characterized by $s$($d$)-wave two-body dominated loss processes correspond to blue circles (squares), while red circles correspond to three-body processes. The green circles indicate loss features in a transitional regime where it is not possible to determine a 2-body or 3-body loss rate. (b) Temperature dependence of the loss rate $\beta(T) \propto T^\alpha$.  (c) Strength of the different loss features. From the results shown in (a) and (b) we derive the two-body and three-body loss rates for a nominal density of $\bar{n} = 1 \times 10^{20} \, \text{m}^{-3}$ and temperature $\bar{T}= \SI{1}{\micro\kelvin}$. (d) Temperature dependence of the two- and three-body loss coefficients $\chi_2 = \alpha +3/2$ (blue) and $\chi_3= \alpha +3$ (red). The 11 loss features studied in this article are marked by vertical bars. 
}
    \label{fig:fig8}
\end{figure}

The same procedure regarding the dependence with density and temperature is applied to the 10 loss features observed in Fig.~\ref{fig:fig1} (top panel) plus the loss feature at \SI{1.755}{\gauss} which emerges for hotter clouds (see Fig.~\ref{fig:fig1} bottom panel). These 11 loss features are identified by the red vertical lines in Fig.~\ref{fig:fig1}. We report in Fig.~\ref{fig:fig8}a the exponent $\gamma$ for each loss feature (with $\beta=-\dot N_a/N_a\propto n_a^\gamma$ at a fixed averaged temperature $\bar{T}$) and in Fig.~\ref{fig:fig8}b the exponent $\alpha$ (with $\beta\propto T^\alpha$ at a fixed initial $N_a$). 

From the results shown in Fig.~\ref{fig:fig8}a, we identify 7 loss features for which the measured value of $\gamma$ is compatible, within error bars, with a two-body dominated process, \textit{i.e.} $\gamma=1$. We identify a single loss feature for which $\gamma$ is compatible, within error bars, with a three-body dominated process, i.e. $\gamma=2$. We also identify three loss features that are not clearly described by either two- or three-body processes. This behaviour can emerge, for example, in the case of the two-step process discussed in \S\,\ref{subsec:dimer_th} when the decay rate of the dimer $\Gamma_d$ is comparable to the rate of the process (\ref{eq:A_plus_A}) producing this dimer. We will not attempt to describe the temperature dependence of those three lines.

Once the assignment of a two-body or three-body resonance has been made from the variation of $\beta$ with $n_a$, we determine the temperature dependence of the associated two- or three-body loss rates, $L_2\propto \beta V$ or $L_3\propto \beta V^2$ for a given $N_a$. 
 Since the volume scales as $T^{3/2}$, we write the temperature dependence of the two- and three-body loss rates as $L _2 \propto T^{\chi{_2}}$ and $L_3 \propto T^{\chi{_3}}$, where $\chi_{2} = \alpha + {3}/{2} $ and $\chi_3 = \alpha +3$.

%%%%%%%%%%%%%%%%%%%%%%%%%%%%%%%%%%%%%%%%%%%%%%
%%%%%%%%%%%%%%%%%%%%%%%%%%%%%%%%%%%%%%%%%%%%%%

In Fig.~\ref{fig:fig8}d we observe that six out of the seven identified two-body processes have a rate $L_2$ compatible with a $1/T$ dependence, as expected for $s$-wave resonances (see Eq.\,(\ref{eq:L2_max}) for $\ell=0$). These resonances are marked with blue circles in the various panels of Fig.\ref{fig:fig8}.

The only exception is the previously unobserved Fano-Feshbach resonance at $B = \SI{1.755}{\gauss}$, marked as a blue square in the various panels of Fig.\ref{fig:fig8}. Its rate scales as $L_2\propto T^{0.5}$. Although the scaling with temperature is strictly speaking not compatible with a $d$-wave resonance for which we would expect $L_2 \propto T$, as observed for chromium \cite{Beaufils2009}, we suggest that the deviation of $\chi_2\approx 0.5$ from 1 is due to the weakness of the loss feature. As shown in Fig.~\ref{fig:fig8}c this resonance is an order of magnitude weaker than the other resonances, and therefore our measurements may be contaminated by other loss processes, such as forced evaporation, which could tend to weaken the observed temperature dependence. 

For the six two-body loss features compatible with a $s$-wave resonance (blue circles in Fig.\ref{fig:fig8}), one might expect a shift of the center of the resonance with temperature, given by $\Delta B = \kb \Delta T/ \delta \mu$, where $\delta \mu$, the differential magnetic moment between open and closed channels, is a priori unknown. 
Given the recent determination of $\delta \mu \approx \SI{1000}{\micro \kelvin / \gauss}$ for the case of thulium \cite{Khlebnikov2021} we would expect a magnetic field shift $\sim \SI{}{\milli\gauss}$, for our temperature range. This shift is comparable to our magnetic field stability and thus difficult to detect on our platform \footnote{We calibrate the magnetic field by transferring a small population in different Zeeman sublevels with a radio-frequency antenna. Due to dipolar relaxation, this process leads to losses when the frequency is resonant with the Zeeman splitting energy $\propto B$. We observed a small drift $\approx \SI{1}{\milli\gauss}$ between different days, which may be due to the stability of our magnetic field or the resolution of the calibration technique.}.

Finally, for the three-body dominated loss feature, identified in red in Fig.~\ref{fig:fig8}, we report a small positive temperature-dependence. At this stage, we cannot conclude whether this feature is caused by a direct three-body resonance as described in \S\ref{subsec:trimer_th}, or whether it corresponds to an effective three-body decay resulting from a particular parameter setting in a two-step two-body loss process as proposed in \cite{Beaufils2009,Li2018,Waseem2018,Waseem2019}.

%%%%%%%%%%%%%%%%%%%%%%%%%%%%%%%%%%%%%%%%%%%%%%
%%%%%%%%%%%%%%%%%%%%%%%%%%%%%%%%%%%%%%%%%%%%%%
%%%%%%%%%%%%%%%%%%%%%%%%%%%%%%%%%%%%%%%%%%%%%%
%%%%%%%%%%%%%%%%%%%%%%%%%%%%%%%%%%%%%%%%%%%%%%

\section{BEC expansion near a loss feature}

Finally, we report a complementary measurement that allows us to determine the $s$-wave scattering length if we suppose that the interactions are of a two-body nature, as it seems to be the case for at least 7 out of 11 resonances.  We perform a long time-of-flight expansion of a Bose-Einstein condensate (BEC) of $^{162}$Dy   and infer the scattering length from its area in the \xyplane~orthogonal to the bias magnetic field pointing along $z$. The BEC is created in a trap with angular frequencies $\px{\omega_x, \, \omega_y,\, \omega_z} = 2\pi \times \px{28,\, 88,\, 66} \SI{}{\hertz}$. We then quench the magnetic field to the desired value and hold the cloud for \SI{20}{\milli\second} before performing a \SI{30}{\milli\second} long time of flight. A magnetic field gradient ensures that the cloud does not fall due to gravity.

The scattering length is derived from a mean-field approach that incorporates dipolar interactions \cite{Eberlein2005}. We write 
\begin{align}
\label{eq:Area}
    a \approx a_\text{stab.} + \pc{\frac{3 R_x R_y}{2}}^{5/2}\times \frac{1}{\mathcal{C} N}
\end{align}
where $a_\text{stab.} \approx 100 \, a_0$ is the scattering length below which the BEC is no longer a stable solution in our trap geometry, $a_0$ is the Bohr radius and $\mathcal{C}\propto a_\text{ho}^4 $ \footnote{Within the Thomas-Fermi approximation, $\mathcal{C} = 15 a_\text{ho}^4 \pc{\frac{\bar{\omega}^2}{\omega_x \omega_y}}^{5/2}$ , where $a_\text{ho} = \sqrt{\frac{\hbar}{m \bar{\omega}}}$ is the harmonic oscillator length. For our trap frequencies we get $\mathcal{C} \approx \SI{31.54}{\micro \meter}^4 $. 
%Magnetic dipolar interactions slightly modify the expression of $\mathcal{C}$, and we numerically find that  $\mathcal{C} \approx \SI{35.881}{\micro \meter}^4$.
}.

%%%%%%%%%%%%%%%%%%%%%%%%%%%%%%%%%%%%%%%%%%%%%%
\begin{figure}[t!]
    \centering
   \includegraphics{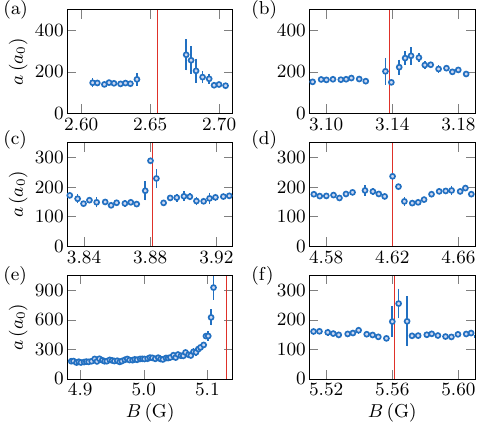}
   \caption{Variation of the scattering length with magnetic field, in the vicinity of six loss features. Central magnetic field:  (a) $B = \SI{2.655}{\gauss}$, (b) $B= \SI{3.138}{\gauss}$, (c) $B=\SI{3.881}{\gauss}$, (d) $B = \SI{4.620}{\gauss}$, (e) $B = \SI{5.130}{\gauss}$, (f) $B = \SI{5.561}{\gauss}$.}
    \label{fig:fig9}
\end{figure}
%%%%%%%%%%%%%%%%%%%%%%%%%%%%%%%%%%%%%%%%%%%%%%

We study the variation of the cloud size for different magnetic fields and show in Fig.~\ref{fig:fig9} the resonances for which a clear dilatation is observed. We interpret this variation as a change in the scattering length using Eq.~\ref{eq:Area}.  Of the 11 loss features studied in this article, only 6 show a clear change in the size of the BEC. These resonances are also the ones with the highest loss rate. Since the measurements are performed with a BEC, we cannot extract negative scattering lengths and report only positive values of $a$. 
We compare our measurements with those reported in Ref.~\cite{Tang2016} and find a good agreement for the scattering length evolution near the $B=\SI{5.130}{\gauss}$ Fano-Feshbach resonance. 

Regarding the only three-body dominated loss feature at $B = \SI{5.561}{\gauss}$ (see Fig.~\ref{fig:fig9}f), we draw the reader's attention to the interpretation of a BEC size change due to a variation of $a$. Since we do not exclude the existence of a pure three-body microscopic process for this resonance, a change in BEC size could also be due to pure three-body interactions and thus not to a change in scattering length. For harmonic traps this implies a change in the total area $\propto N^{1/2}$ (instead of $N^{2/5}$ for two-body interactions). We have tested this hypothesis by varying the number of atoms in the BEC. Although our results are consistent with $N^{2/5}$ scaling, we cannot exclude the $N^{1/2}$ result within our experimental uncertainties. It will be particularly interesting to study this resonance in a flat-bottom trap, where the size of the BEC area after time-of-flight will be proportional to $N^3$ for three-body interactions instead of $N^2$ for two-body interactions, leading to an easier lift of ambiguity. However, this study is beyond the scope of this article.

%%%%%%%%%%%%%%%%%%%%%%%%%%%%%%%%%%%%%%%%%%%%%%
%%%%%%%%%%%%%%%%%%%%%%%%%%%%%%%%%%%%%%%%%%%%%%
%%%%%%%%%%%%%%%%%%%%%%%%%%%%%%%%%%%%%%%%%%%%%%
%%%%%%%%%%%%%%%%%%%%%%%%%%%%%%%%%%%%%%%%%%%%%%

\section{Conclusion}

We have characterized 11 low-field resonant loss features of an ultracold thermal sample of dysprosium. From the analysis of their density and temperature dependence, we conclude that most loss features result from a resonant $s$-wave pairwise interaction. We also measured the corresponding change in scattering length through the expansion of a BEC. Additionally, we measured the three-body loss rate of a quasi-pure BEC near zero magnetic field, which provides valuable information for future studies aimed at exploiting spin-orbit coupling in a dense condensate of dysprosium atoms \cite{Chalopin2020}.

Furthermore, we have evidenced (i) a possible two-body $d$-wave resonance, unobserved so far, highlighting the complex nature of interactions in strongly dipolar gases, and  (ii) a loss resonance that may be controlled by pure three-body interactions. To go further in our quantitative description of three-body resonances, whether driven by two-body or three-body processes, we will need to better disentangle the roles of temperature and density, which should be possible with the preparation of a homogeneous sample of dysprosium in a box-like trap \cite{Navon2021}.

\textit{Acknowledgements: } We thank Sarah Wattellier, Jean-Gabriel Pipelin, Louis Chambard, Virgin Durepaire, and Julie Veschambre for early contributions to the experimental setup. We acknowledge fruitful discussions with the members of the Bose-Einstein condensate team at LKB, and with F. Werner, D. Petrov, F. Chevy, R. P. Smith, O. Dulieu and M. Lepers. This work was supported by Grant No. ANR-20-CE30-0024, and by Region Ile-de-France in the framework of DIM QuanTiP.

\appendix*
\section{Details of the experimental setup}

%%%%%%%%%%%%%%%%%%%%%%%%%%%%%%%%%%%%%%%%%%%%%%
%%%%%%%%%%%%%%%%%%%%%%%%%%%%%%%%%%%%%%%%%%%%%%
%%%%%%%%%%%%%%%%%%%%%%%%%%%%%%%%%%%%%%%%%%%%%%
%%%%%%%%%%%%%%%%%%%%%%%%%%%%%%%%%%%%%%%%%%%%%%

\subsection{Production of a $^{162}$Dy  MOT }

%%%%%%%%%%%%%%%%%%%%%%%%%%%%%%%%%%%%%%%%%%%%%%
\begin{figure}[t!]
    \centering
   \includegraphics[width=0.5\textwidth]{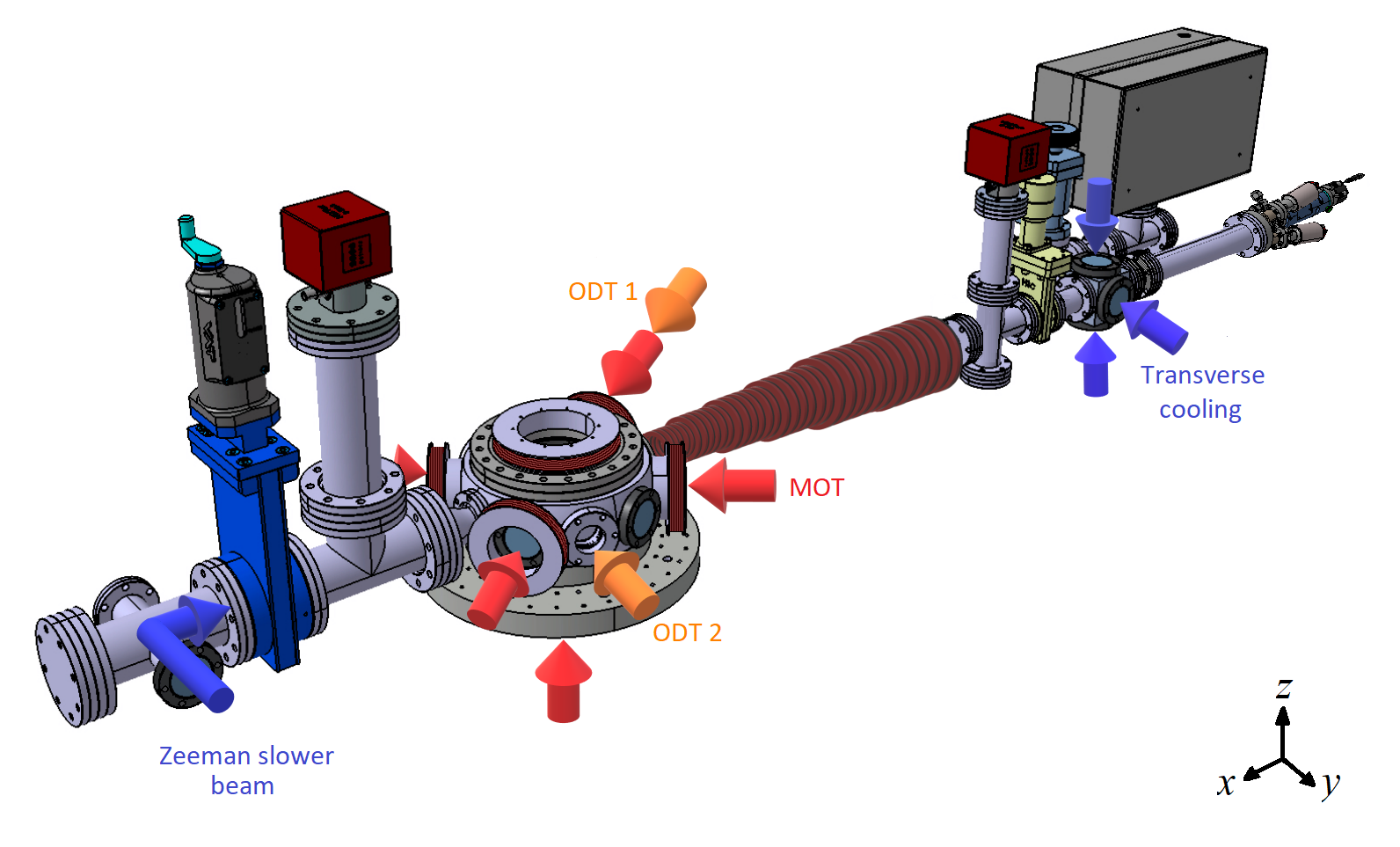}
   \caption{Schematic representation of the experimental setup. The atoms are optically collimated at the output of the oven (transverse cooling) and then decelerated in a spin-flip Zeeman slower. The atoms are then confined in a magneto-optical trap (MOT), which comprises 5 laser beams and a magnetic field gradient. We compress the MOT and capture approximately $4\times 10^6$ atoms into a crossed dipole trap (CDT) made up of two beams (ODT1 and ODT2) with a relative angle of $144 ^\circ$.}
    \label{fig:figmanip}
\end{figure}
%%%%%%%%%%%%%%%%%%%%%%%%%%%%%%%%%%%%%%%%%%%%%%

We load 3$\times 10^8$ atoms in a magneto-optical-trap (MOT), with a loading rate of $1 \times 10^8 \, \text{atom}\SI{}{/\second}$, composed of five red-detuned beams (four in the \xyplane~and one vertical), with detuning $\Delta_\text{MOT} = -43 \times {\Gamma_\text{red}}$, with respect to the intercombination line with vacuum wavelength $\lambda = \SI{626.08}{\nano\meter}$ (red transition) and linewidth $\Gamma_\text{red} \approx 2\pi \times \SI{136}{\kilo\hertz}$ \cite{Ilzhofer2018}. The MOT is produced with a saturation parameter $s_0= I_0 / I_\text{sat.} \approx 185$ (the vertical beam has a tenth of this intensity) and a magnetic field gradient $\partial_z B_z = \SI{1.72}{\gauss/\centi\meter}$. The cloud lies a few \SI{}{\milli\meter} below the zero-field, which makes it fully polarized in the Zeeman sublevel of lowest energy $\ket{J, m_J =-J}$.  The MOT capture velocity, determined by the linewidth of the transition and the MOT beam waist, $w_\text{MOT} = \SI{15}{\milli\meter}$, is equal to $v_c \approx \SI{7}{\meter/\second}$. We maximize the atomic flux reaching the main science chamber using an optical collimation. It is achieved through the cooling provided by a transverse molasses consisting of two retroreflected laser beams with orthogonal propagation with respect to the oven's exit axis. 
The molasses laser beams are red-detuned, with detuning $\Delta_\text{Coll.} = -0.4\times{\Gamma_\text{blue}}$,  with respect to the optical transition with vacuum wavelength of \SI{421.29}{\nano\meter} (blue transition) and linewidth $\Gamma_\text{blue} \approx 2\pi \times\SI{32}{\mega\hertz}$. A saturation parameter of $s_0= 4$ is used, leading to a five-fold increase in the loading rate of the MOT. The longitudinal velocity of the atomic flux exiting the oven is reduced from $v_x \approx \SI{400}{\meter/\second}$ to $\approx \SI{7}{\meter/\second}$ in a \SI{500}{\milli\meter} Zeeman slower in a spin-flip configuration, with a laser beam detuning of $\Delta_\text{Zeeman} = -14\times{\Gamma_\text{blue}}$ and a saturation parameter of $s_0 \approx 1$. 
The blue and red laser beams are frequency stabilized through modulation frequency transfer using either the atomic flux at the output of the oven (for the blue transition), or an iodine cell  (for the red transition)~\cite{Lucioni2017}.

%%%%%%%%%%%%%%%%%%%%%%%%%%%%%%%%%%%%%%%%%%%%%%
\begin{figure}[b!]
    \centering
   \hspace{-0.4cm}\includegraphics[width=0.5\textwidth]{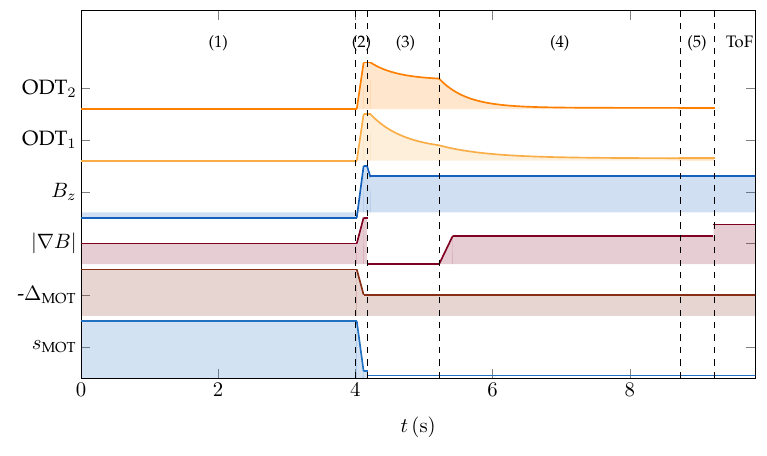}
   \caption{Schematic representation of the experimental sequence used to produce a degenerate gas of dysprosium. (1) MOT loading stage. (2) Compressed MOT. (3) First stage of evaporative cooling. (4) Second stage of evaporative cooling. (5) Plain evaporation in the CDT to purify the quantum gas. (ToF) Time-of-flight expansion and absorption imaging acquisition.}
    \label{fig:sequence}
\end{figure}
%%%%%%%%%%%%%%%%%%%%%%%%%%%%%%%%%%%%%%%%%%%%%%

\subsection{Compressed MOT and transfer into a crossed dipole trap}
\label{sec:app2}

To further cool the atomic cloud, we reduce the saturation parameter of the red lasers to $s_0 = 5.7$, decrease the detuning to $\Delta_{cMOT} = -21\times \Gamma_\text{red}$, and increase the gradient to $\partial_z B_z = \SI{4.31}{\gauss/\centi\meter}$, which stabilizes the position of the cloud against small fluctuations in the laser frequency and magnetic field. Although it leads to a compressed MOT (cMOT) with a temperature $T\approx \SI{15}{\micro\kelvin}$, higher than the Doppler temperature, it ensures a stable overlap with the crossed dipole trap (CDT). The cMOT has a typical size at $1/\sqrt{e}$ of \SI{400}{\micro\meter} and atom number $\sim 1\times10^8$.

%%%%%%%%%%%%%%%%%%%%%%%%%%%%%%%%%%%%%%%%%%%%%%
\begin{figure}[t!]
    \centering
   \includegraphics[width=0.5\textwidth]{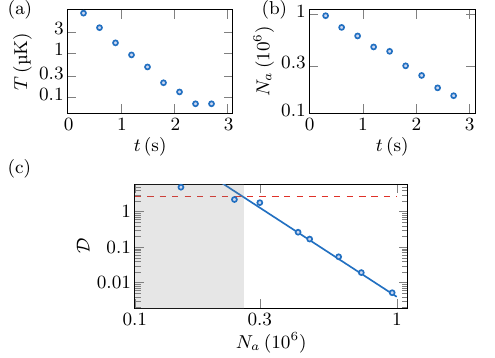}
   \caption{Efficiency of evaporative cooling. (a) Temperature as a function of time in log-linear scale. (b) Atom number as a function of time in log-linear scale. (c) Phase-space-density ($\mathcal{D}$) as a function of atom number, $N_a$, in log-log scale. We fit our data with  $\mathcal{D} \propto N_a^{-\vartheta}$ and retrieve $\vartheta \approx 4.0$. The gray region indicates the points for which a non-negligible condensed fraction is already present and the estimation of $\mathcal{D}$ is no longer quantitative. The horizontal dashed line corresponds to the non-interacting prediction for the emergence of a BEC at $\mathcal{D} \approx 2.612$.}
    \label{fig:figPSD}
\end{figure}
%%%%%%%%%%%%%%%%%%%%%%%%%%%%%%%%%%%%%%%%%%%%%%

The CDT is composed of two single-mode laser beams at a wavelength of \SI{1064}{\nano\meter}, with a relative angle of $144^\circ$ in the \xyplane, and a frequency difference of \SI{180}{\mega\hertz}, which ensures that residual interference patterns are averaged out. As we reduce the detuning of the MOT beams, light-induced loss processes are enhanced, resulting in a short lifetime of the cMOT ($\sim \SI{200}{\milli\second}$). The loading of the cMOT into the CDT is therefore fast $\sim \SI{50}{\milli\second}$, which allows us to load $4\times 10^6$ atoms at a temperature of \SI{60}{\micro\kelvin}. The two beams that make up the CDT, hereafter identified as ODT1 and ODT2 (see Fig.~\ref{fig:figmanip}), have waists $w_\text{ODT1} \approx \SI{30}{\micro\meter}$ and $w_\text{ODT2} \approx \SI{20}{\micro\meter}$. 
To optimize the loading, we enlarge the waist of ODT1 in the \xyplane~by a factor of 2 with an acousto-optic deflector \cite{Henderson2009}. The two optical dipole trap beams have maximum powers of \SI{30}{\watt} (ODT1) and \SI{5}{\watt} (ODT2) and the forced evaporative cooling is performed in the presence of a bias field aligned along the $z$-direction and with magnitude $B_z = \SI{1.660}{\gauss}$.
The maximum loading efficiency is achieved for horizontally polarized dipole beams. This is because the difference in polarizability between the ground and excited states of the intercombination line, $\Delta \alpha = \alpha_\text{exc.} - \alpha_\text{ground}$, is negative at \SI{1064}{\nano\meter} for a linear horizontal polarization \cite{Chalopin2018}. This guarantees that the detuning of the cMOT cooling beam remains negative. After loading the CDT and prior to evaporative cooling, we rotate the polarization of ODT2 by 90$^\circ$.

%%%%%%%%%%%%%%%%%%%%%%%%%%%%%%%%%%%%%%%%%%%%%%
%%%%%%%%%%%%%%%%%%%%%%%%%%%%%%%%%%%%%%%%%%%%%%
%%%%%%%%%%%%%%%%%%%%%%%%%%%%%%%%%%%%%%%%%%%%%%
%%%%%%%%%%%%%%%%%%%%%%%%%%%%%%%%%%%%%%%%%%%%%%

\subsection{Crossed dipole trap evaporative cooling}

The first evaporation stage lasts for \SI{1}{\second}, during which we suppress the spatial modulation of ODT1 and reduce its power so that the two beams forming the crossed dipole trap have similar trap depths. We focus here on the second stage of forced evaporative cooling, which lasts for \SI{3.5}{\second} (see Fig.~\ref{fig:sequence}). At the beginning of this evaporation, we switch on a magnetic field gradient to partially compensate for gravity, which plays an important role in the optimization of the evaporative cooling \cite{Hung2008}.

%%%%%%%%%%%%%%%%%%%%%%%%%%%%%%%%%%%%%%%%%%%%%%

\begin{figure}[t!]
    \centering
   \includegraphics{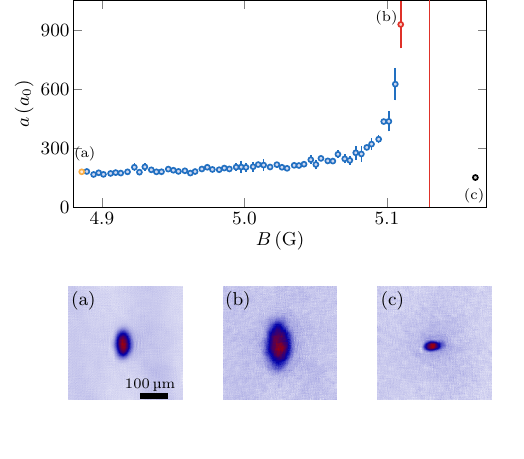}
  \caption{Scattering length in the vicinity of the $B_0 = \SI{5.130}{\gauss}$ resonance (vertical red line). 
  Expansion of a BEC in the $x$--$y$~plane after \SI{30}{\milli\second} time-of-flight, (a) far from the Feshbach resonance and near the resonance with $B<B_0$ (b) 
  or $B>B_0$ (c). We interpret the strong expansion of the BEC for $B<B_0$ resulting from a large scattering length, while for $B>B_0$ we observe a dense cloud compatible with the formation of a quantum droplet.}
    \label{fig:expansion}
\end{figure}
%%%%%%%%%%%%%%%%%%%%%%%%%%%%%%%%%%%%%%%%%%%%%%

We measure both temperature, atom number and trap frequencies at different times during the second evaporative cooling stage. In Fig.~\ref{fig:figPSD}a, b we show the evolution of temperature and atom number as a function of time. Combined with our measurements of the trap frequencies, we compute the evolution of the averaged phase-space density $\mathcal{D} = n_a \lambda^3$, where $n_a$ takes into account the averaging over the inhomogeneous density in the harmonic potential (see main text). We show in Fig.~\ref{fig:figPSD}c the evolution of $\mathcal{D}$ as a function of $N_a$ in log-log scale. From the fit function $\mathcal{D} \propto N_a^{-\vartheta}$, we determine the evaporative cooling efficiency $\vartheta \approx 4$, at the upper end of typical values in optical dipole traps ($2\lesssim \vartheta \lesssim 5$)  \cite{Hung2008}.\\

%%%%%%%%%%%%%%%%%%%%%%%%%%%%%%%%%%%%%%%%%%%%%%
%%%%%%%%%%%%%%%%%%%%%%%%%%%%%%%%%%%%%%%%%%%%%%
%%%%%%%%%%%%%%%%%%%%%%%%%%%%%%%%%%%%%%%%%%%%%%
%%%%%%%%%%%%%%%%%%%%%%%%%%%%%%%%%%%%%%%%%%%%%%
%\subsection{Two-photon heating and loss in a multimode \SI{1070}{\nano\meter} IPG laser}
% 
%%%%%%%%%%%%%%%%%%%%%%%%%%%%%%%%%%%%%%%%%%%%%%
%%%%%%%%%%%%%%%%%%%%%%%%%%%%%%%%%%%%%%%%%%%%%%
%%%%%%%%%%%%%%%%%%%%%%%%%%%%%%%%%%%%%%%%%%%%%%
%%%%%%%%%%%%%%%%%%%%%%%%%%%%%%%%%%%%%%%%%%%%%%

\appendix*
\section{Examples of expanded clouds in the vicinity of the \SI{5.130}{\gauss} Feshbach resonance}

We here present some examples for the BEC expansion near the $B_0 =\SI{5.130}{\gauss}$ Feshbach resonance discussed in Fig.~\ref{fig:fig9}(e). We deduce the change in scattering length from the cloud dilatation in the \xyplane~after a \SI{30}{\milli\second} time-of-flight. As shown in Fig.~\ref{fig:expansion}a, the expansion of the BEC far from the resonance, for instance $B = \SI{4.886}{\gauss}$ (yellow dot), leads to radii of \SI{30}{} - \SI{45}{\micro\meter}, along the two orthogonal axes of the optical dipole trap in the \xyplane. As the scattering length increases, the cloud expands significantly more, as shown in Fig.~\ref{fig:expansion}b for a magnetic field of $B = \SI{5.109}{\gauss}$. Interestingly, for $B>B_0$, where the scattering length passes from large negative values to its background value, and thus crosses zero, we observe a cloud that does not expand even after a long time-of-flight (see Fig.~\ref{fig:expansion}c), which is a hallmark of quantum droplets \cite{FerrierBarbut2016}. 
Our procedure for determining the scattering length cannot be used in this case, as one would need to incorporate beyond-mean-field corrections to explain the stability of the cloud.

\bibliographystyle{ieeetr}
\bibliography{bib.bib}

\end{document}